\documentstyle[12pt]{article}
\begin{document}
\begin{flushright}
\vspace{1mm}
 FIAN/TD/04--97\\
{March 1997}\\
\end{flushright}\vspace{2cm}

\begin{center}
{\Large Lagrangian formulation of irreducible massive fields\\
of arbitrary spin in 2+1 dimensions}
\end{center}

\centerline{I.V.Tyutin and M.A.Vasiliev}

\begin{center}
I.E.~Tamm Theory Department, P.N.~Lebedev Physical Institute,
117924,\\ Leninsky prospect 53, Moscow, Russia.
\end{center}

\vskip1.3cm

Lagrangian formulation of free massive fields corresponding
to irreducible representations of the Poincare group of arbitrary
integer and half--integer spins in three--dimensional space--time
is presented.
A relationship of the theory under consideration with the self--duality
equations in four dimensions is discussed.

\vskip2cm

\section{Introduction}

A purpose of the present paper is to give a Lagrangian formulation
of free massive fields corresponding to an irreducible representation
of the Poincare group of an arbitrary integer (bosons) or half--integer
(fermions) spin in the three-dimensional
space-time. An interest to
lower dimensional models is due to the fact that these relatively
simple theories serve as a polygon for study of more realistic models in
higher-dimensional spaces. Also one can hope that such models can
describe  real film--type physical systems.

After the classical work by Fierz and Pauli
\cite{FP}, there was a great number of papers
in which gauge invariant actions were formulated for
massless fields of an arbitrary spin in any dimension
\cite{*} as well as actions for massive fields corresponding
to totally symmetric representations of the little group in
dimensions $D\ge4$ \cite{**}. In the work
\cite{ADY}, it was demonstrated that the actions for massive
fields in  $D$ dimensions can be derived by means of the
dimensional reduction of the actions for massless fields
in $D+1$ dimensions. This automatically gives rise to
supplementary fields necessary for the formulation of a local action.
A straightforward application of such a procedure to a derivation of
the actions of massive fields in 3 dimensions via reduction of the
gauge invariant action in 4
dimensions
\cite{F,FF,DF,v1,ad} faces however the following problem.
Gauge invariant actions in 4 dimensions describe fields possessing
two helicities (these fields can be interpreted either as Majorana
fields with two values of helicity or as Weyl fields and antifields,
each having a fixed value of helicity).
Dimensional reduction to the three--dimensional space gives
rise to an action which describes a field with two values
of the
(2+1)--dimensional spin, which therefore constitutes
some reducible representation of the
(2+1)--dimensional Poincare group \cite{B}.
Since it is impossible to build a local action in four dimensions
which describes a self--dual field possessing a unique value
of helicity \cite{Sh} (i.e. Majorana-Weyl field),
after dimensional reduction
 4 $\rightarrow$ 3 it is still necessary to carry out an additional
reduction which leaves a half of physical degrees of freedom if one
wishes to have a physical system which corresponds to an irreducible
representation of the Poincare group in 2+1 dimensions.

So far the actions for irreducible massive fields in (2+1)--dimensional
space with spins
$s=1$ \cite{T,TOP1,DJT}, $s=3/2$ \cite{DK,D}, $s=2$ \cite{DJT,AK} and s=5/2
\cite{AS,AD} have been constructed.
The formalism which we apply in this paper for the Lagrangian
description of arbitrary irreducible representations of (2+1)--
dimensional Poincare group is related to the
reduction to 2+1 dimensions of the vierbein-type variables introduced
for the description of massless fields of an arbitrary integer
\cite{v1} and half-integer
\cite{v1,ad} spin in  3+1 dimensions.  For the cases of spins 3/2, 2 and
 5/2 we reproduce the results obtained in
\cite{D}, \cite{DJT} and  \cite{AS}, respectively.

A structure of the paper is as follows.
In the section 2 we summarize the main notations used throughout the
paper. In the section 3 the actions, which are of first order in derivatives,
are constructed for boson fields of an arbitrary integer spin
 $s$, $|s|\ge2$. In the section 4 an analogous problem is solved for
fermion fields of an arbitrary half-integer spin
$s$, $|s|\ge3/2$. The conditions on the parameters of the model
(i.e. sign choices in front of the kinetic and mass terms in the action)
which follow  from the requirements of boundness of energy and positive
definiteness of norms of states are derived in the section 5.
 In section 6, an  action is presented
for the bosonic case of integer spin $s$,
which is of the second order
in derivatives and
describes
the both irreducible representations of the
(2+1)--dimensional Poincare group for a fixed value of $|s|$.

\section{Notations and Problem Setting}

Our field variables include the fields
$\lambda^n_{\alpha(n)}\equiv\lambda^n_{\alpha_1\ldots\alpha_n}$ and
$h^n_{\mu|\alpha(n)}\equiv h^n_{\mu|\alpha_1\ldots\alpha_n}$,
which are totally symmetric in spinor indices
($\mu,\nu,\ldots=0,1,2,\quad$
$\alpha,\beta,\ldots=1,2$)\footnote{Greek indices from the first part
 of the Alphabet are spinorial while Greek indices from the second part
of the alphabet are vectorial.}.
Also we will use the shorthand notations
$\lambda^n$ and $h^n_\mu$ for these fields.
Vectorial indices are raised and lowered as usual by virtue of
the metric tensors
$\eta^{\mu\nu}=\eta_{\mu\nu}=\hbox{diag}(+,-,-)$.
Spinorial indices are raised and lowered by virtue of
$\varepsilon_{\alpha\beta} =-\varepsilon_{\beta\alpha}$
and
$\varepsilon^{\alpha\beta}= -\varepsilon^{\beta\alpha}$,
$\varepsilon_{12}=\varepsilon^{12}=1$:

$$
A_\alpha=A^\beta\varepsilon_{\beta\alpha},\quad
A^\alpha=\varepsilon^{\alpha\beta}A_\beta,\quad
A^\alpha B_\alpha=-A_\alpha B^\alpha.
$$

The Lorentz transformations of fields
$h^n_\mu$
and
$\lambda^n$,
which correspond to the transformation of the coordinates
$\delta x^\mu=
\omega^{\mu\nu}x_\nu$, $\omega^{\mu\nu}=-\omega^{\nu\mu}$,
have a form:
$$
\delta h^n_{\mu|\alpha(n)}={i\over2}\omega^{\nu\sigma}
(M_{\nu\sigma}h^n)_{\mu|\alpha(n)}=
$$
\begin{equation}\label{1}
\omega^{\nu\sigma}x_\nu\partial_\sigma h^n_{\mu|\alpha(n)}+
\omega_{\mu\nu}h^{n\nu|}{}_{\alpha(n)}-{n\over4}\omega^{\nu\sigma}
\varepsilon_{\nu\sigma\rho}e^{\rho|}{}_\alpha{}^\beta
h^n_{\mu|\alpha(n-1)\beta},
\end{equation}
\begin{equation}\label{2}
\delta \lambda^n_{\alpha(n)}=
\omega^{\nu\sigma}x_\nu\partial_\sigma \lambda^n_{\alpha(n)}-
{n\over4}\omega^{\nu\sigma}\varepsilon_{\nu\sigma\rho}
e^{\rho|}{}_\alpha{}^\beta\lambda^n_{\alpha(n-1)\beta}.
\end{equation}
Here $\partial_\mu\equiv\partial/\partial x^\mu$,
$\varepsilon_{\mu\nu\sigma}$ is the totally antisymmetric tensor
$\varepsilon_{012}=\varepsilon^{012}=1$, while
$e^{\mu|}{}_{\alpha\beta}$ is a set of three real symmetric matrices
which are defined by the relation
$$
e^{\mu|}{}_{\alpha\gamma}e^{\nu|}{}_\beta{}^\gamma=
\eta^{\mu\nu}\varepsilon_{\alpha\beta}-\varepsilon^{\mu\nu\sigma}
e_{\sigma|\alpha\beta}
$$
and satisfy the following useful identities:
$$
e_{\mu|\alpha(2)}e^{\mu|\beta(2)}=2\delta_\alpha^\beta\delta_\alpha^\beta,
\quad
\varepsilon^{\mu\nu\sigma}e_{\nu|\alpha(2)}e_{\sigma|}{}^{\beta(2)}=
-2\delta_\alpha^\beta e^{\mu|}{}_\alpha{}^\beta
$$
(in the section 6 we will use  a special realization with
$e^{0|}{}_{\alpha\beta}=\delta_{\alpha\beta}$).
We use the following convention
\cite{v0} :
$$
A^{k+l}_{\alpha(k)\beta(l)}B^{m+n}_{\alpha(m)\gamma(n)}\equiv
{1\over C^k_{k+m}}\sum A^{k+l}_{\alpha_{i_1}\ldots\alpha_{i_k}\beta(l)}
B^{m+n}_{\alpha_{i_{k+1}}\ldots\alpha_{i_{k+m}}\gamma(n)},
$$
where the summation is carried out over all
$C^k_{k+m}$ combinations
of indices
$\alpha_i$ ($A^{k+l}$
and
$B^{m+n}$ are symmetric with respect
to their spinorial indices.)
Also
$$
A^{n+m\alpha(n)\beta(m)}B^{n+k+l}_{\alpha(n+k)\gamma(l)}\equiv\sum_{\alpha_i}
A^{n+m\alpha_1\ldots\alpha_n\beta(m)}
B^{n+k+l}_{\alpha_1\ldots\alpha_n\alpha(k)\gamma(l)},
$$
$$
A^{n+m\alpha(n)\beta(m)}B^{n_1+k}_{\alpha(n_1)\gamma(k)}
C^{n-n_1+l}_{\alpha(n-n_1)\delta(l)}=
A^{n+m\alpha(n)\beta(m)}\left(B^{n_1+k}_{\alpha(n_1)\gamma(k)}
C^{n-n_1+l}_{\alpha(n-n_1)\delta(l)}\right).
$$

The Casimir operator of the Poincare group (the Pauli--Lubanski
scalar) $W$,
$$
W=-{i\over2}\varepsilon^{\mu\nu\sigma}M_{\nu\sigma}\partial_\mu,
$$
acts on $h^n_\mu$ according to the rule:
\begin{equation}\label{3}
(Wh^n)_{\mu|\alpha(n)}=
\varepsilon_\mu{}^{\nu\sigma}\partial_\nu h^n_{\sigma|\alpha(n)}+
{n\over2}e^{\sigma|}{}_\alpha{}^\beta\partial_\sigma
h^n_{\mu|\alpha(n-1)\beta}.
\end{equation}
Its eigenvalues
$ms$ determine a spin $s$  of a state with respect of the Poincare
group
while $m$ is a mass of a state which will be always assumed to be
positive.

A field $h^N_\mu$ describes a single (highest) spin
$\pm(N+2)/2$ if it satisfies the conditions,
\begin{equation}\label{4}
e^{\mu|}{}_\beta{}^\gamma
h^N_{\mu|\alpha(N-1)\gamma}=0
\end{equation}
(which implies that the field
$\tilde{h}^{N+2}_{\beta_1\beta_2\alpha_1\ldots\alpha_N}\equiv
(1/2)e^{\mu|}{}_{\beta_1\beta_2}h^N_{\mu|\alpha_1\ldots\alpha_N}$
is symmetric with respect to all spinor indices) and, e.g.
\begin{equation}\label{5}
\varepsilon_\mu{}^{\nu\sigma}\partial_\nu
h^N_{\sigma|\alpha(N)}=\pm mh^N_{\mu|\alpha(N)}
\end{equation}
which implies that the field
$\tilde h^{N+2}$ satisfies a Dirac equation with respect to each
spinorial index.

Our goal is to derive an extremal principle
which gives rise to the equations
(\ref{4}), (\ref{5}) and, may be,  to some dynamically trivial
equations for supplementary fields.

\section{Boson Fields}

In this section we construct an action for boson fields of an arbitrary
integer spin
$s$, $|s|=(N+2)/2\ge2$, $N\ge2$ -- even.
Let us start with the action:
$$
S_{B}=
{1\over2}\sum_{n=0}^N\xi_n\varepsilon^{\mu\nu\sigma}h^n_{\mu|}{}^{\alpha(n)}
\partial_\nu h^n_{\sigma|\alpha(n)}+{m\over2}\sum_{n=2}^N
a_n\varepsilon^{\mu\nu\sigma}h^n_{\mu|}{}^{\alpha(n-1)\beta}
e_{\nu|\beta}{}^\gamma h^n_{\sigma|\alpha(n-1)\gamma}+
$$
\begin{equation}\label{6}
+m\sum_{n=2}^N b_n\varepsilon^{\mu\nu\sigma}h^n_{\mu|}{}^{\alpha(n)}
e_{\nu|\alpha(2)} h^{n-2}_{\sigma|\alpha(n-2)}+
m\sum_{n=0}^{N-2}\lambda^{n+2}_{\alpha(n+2)}
e^{\mu|\alpha(2)}h^n_{\mu|}{}^{\alpha(n)},
\end{equation}
where summation is carried out over even
$n$. Boson fields
$h^n_\mu$, $\lambda^n$  are real. {}From the requirement that
the action has to be real it follows that the coefficients
$\xi_n$, $a_n$, $b_n$ should be real.
We assume that
$\xi_n=\pm1$ that can always be achieved by an appropriate
rescaling of fields.
The action
(\ref{6}) is invariant under the Lorentz transformations
(\ref{1}), (\ref{2}).
The problem consists of choosing such coefficients
$\xi_n$, $a_n$, $b_n$ that the extra coefficients of the field
 $h^N_\mu$ (including one of the highest spins)  vanish as
a result of field equations, as well as all fields
$h^n_\mu$ with $0\le n\le N-2$, and $\lambda^n$, $2\le n\le N$.

The variation of the action with respect to
$\lambda^n$
gives the constraints
\begin{equation}\label{8}
e^{\mu|}{}_{\alpha(2)}h^n_{\mu|\alpha(n)}=0,\quad 0\le n\le N-2.
\end{equation}
{}From (\ref{8}) at $n=0$ it follows that
\begin{equation}\label{7}
h^0_\mu=0.
\end{equation}

The variation of the action over
$h^n_\mu$ gives the equations of motion:
$$
\xi_n\varepsilon^{\mu\nu\sigma}\partial_\nu h^n_{\sigma|\alpha(n)}+(1-
\delta_{n,0})ma_n\varepsilon^{\mu\nu\sigma}e_{\nu|\alpha}{}^\beta
h^n_{\sigma|\alpha(n-1)\beta}+(1-\delta_{n,0})mb_n\varepsilon^{\mu\nu\sigma}
e_{\nu|\alpha(2)}h^{n-2}_{\sigma|\alpha(n-2)}-
$$
\begin{equation}\label{9}
(1-\delta_{n,N})mb_{n+2}\varepsilon^{\mu\nu\sigma}e_{\nu|}{}^{\alpha(2)}
h^{n+2}_{\sigma|\alpha(n+2)}+(1-\delta_{n,N})me^{\mu|\alpha(2)}
\lambda^{n+2}_{\alpha(n+2)}=0.
\end{equation}

Let us contract the equations
(\ref{9}) with $\partial_\mu$ and exclude derivatives of all
functions
$h^n_\sigma$ in the resulting equations with the aid of
(\ref{9}). For every
$n$ one is then left with some terms containing
$h^{n+4}_\sigma$,
$h^{n+2}_\sigma$, $h^n_\sigma$, $h^{n-2}_\sigma$ and $h^{n-4}_\sigma$.
The coefficients in front of
$h^{n+4}_\sigma$ and $h^{n-4}_\sigma$ vanish identically
due to the properties of the matrices
$e_\mu$ and the antisymmetrization over vectorial indices.
A term with
$h^{n-2}_\sigma$ vanishes as a consequence of the constraints
(\ref{8}).\footnote{Note that at this point only one of the
constraints is really necessary, namely
(\ref{7}). For  $n\ge4$ the coefficients in front of
$h^{n-2}_\sigma$ vanish provided that
 $a_n$ have a form
(\ref{12}).} As a result one derives after some transformations:
\begin{equation}\label{10}
2\xi_2a_2b_2e^{\mu|\alpha(2)}h^2_{\mu|\alpha(2)}={1\over m}
e^{\mu|\alpha(2)}\partial_\mu\lambda^2_{\alpha(2)},
\end{equation}
$$
b_{n+2}\left(\xi_na_n-{n+4\over
n+2}\xi_{n+2}a_{n+2}\right)e^{\sigma|\alpha(2)}
h^{n+2}_{\sigma|\alpha(n+2)}+
$$
$$
\left({2\over n}\xi_na^2_n+2\xi_{n-2}b^2_n-
{2n(n+3)\over(n+1)(n+2)}\xi_{n+2}b^2_{n+2}\right)e^{\sigma|}{}_\alpha{}^\beta
h^n_{\sigma|\alpha(n-1)\beta}+
$$
\begin{equation}\label{11}
2\xi_{n-2}b_n\lambda^n_{\alpha(n)}+
{1\over m}(1-\delta_{n,N})e^{\mu|\alpha(2)}\partial_\mu
\lambda^{n+2}_{\alpha(n+2)}=0,
\quad 2\le n\le N,
\end{equation}
where, by definition,
$b_{N+2}=0$.

Let us require the coefficients in front of
$h^{n+2}_\sigma$ and $h^n_\sigma$ in (\ref{11}) to vanish:
$$
\xi_na_n-{n+4\over n+2}\xi_{n+2}a_{n+2}=0,\quad 2\le n\le N,
$$
$$
{1\over n}a^2_n+\xi_{n-2}b^2_n-{n(n+3)\over(n+1)(n+2)}\xi_{n+2}b^2_{n+2}=0,
\quad 2\le n\le N.
$$
A solution of these equations reads:
$$
a_n={N+2\over n+2}\xi_N\xi_n(-1)^{(N+2)/2}\theta,\quad
b^2_n=-\xi_{n-2}\xi_n{(N+2)^2-n^2\over4n(n+1)},\quad \theta^2=1,
$$
where we have chosen a normalization
$a_N=(-1)^{(N+2)/2}\theta$, which can always be achieved by
virtue of some redefinition of the parameter
$m$. In the section 5 it will be shown that the requirements of
boundness of energy and positive definiteness of the norms of
 states is satisfied only for
$\theta=1$ which condition is assumed to be true from now on. The condition
$b^2_n>0$ gives
$\xi_{n-2}\xi_n=-1$, i.e.
$$
\xi_n=(-1)^{(n+2)/2}\xi, \quad \xi^2=1,
$$
so that one finally gets
\begin{equation}\label{12}
\xi_n=(-1)^{(n+2)/2}\xi,\quad a_n=(-1)^{(n+2)/2}{N+2\over n+2},\quad
b_n={(-1)^{(n+2)/2}
\over2}\sqrt{{(N+2)^2-n^2\over n(n+1)}}.
\end{equation}
Any other choice of signs for
$b_n$ can be compensated by a simple field redefinition.

Setting
$n=N$ in (\ref{11}), we obtain

\begin{equation}\label{11'}
\lambda^N_{\alpha(N)}=0.
\end{equation}

Setting then $n=N-2$ in (\ref{11}) and taking into account (\ref{11'}),
one gets
$$
\lambda^{N-2}_{\alpha(N-2)}=0,
$$
{\it etc}.

Finally, from (\ref{11}) it follows that all Lagrange multipliers
vanish as a consequence of the equations of motion
$$
\lambda^n_{\alpha(n)}=0, \quad n=2,\ldots,N.
$$

Let us now analyze the equations for
$h^n_\mu$.

Let $N=2$.

The equation of motion
(\ref{9}) at $n=0$ (taking into account (\ref{7})) along with the equation
(\ref{10}) give
$$
\varepsilon^{\mu\nu\sigma}e_{\nu|}{}^{\alpha(2)}h^2_{\sigma|\alpha(2)}=0,\quad
e^{\mu|\alpha(2)}h^2_{\mu|\alpha(2)}=0,
$$
that is equivalent to the condition
(\ref{4}). The equation of motion
(\ref{9}) at $n=2$ has a form
$$
\varepsilon^{\mu\nu\sigma}\partial_\nu h^2_{\sigma|\alpha(2)}+
\xi m\varepsilon^{\mu\nu\sigma}e_{\nu|\alpha}{}^\beta
h^2_{\sigma|\alpha\beta}=0,
$$
which is equivalent to (\ref{5}).

Let $N>2$.

The equation of motion
(\ref{9}) at $n=0$ (taking into account (\ref{7})), the constraint
(\ref{8}) at $n=2$ and the equation
 (\ref{10}) give
$$
\varepsilon^{\mu\nu\sigma}e_{\nu|}{}^{\alpha(2)}h^2_{\sigma|\alpha(2)}=0,\quad
e^{\mu|}{}_{\alpha(2)}h^2_{\mu|\alpha(2)}=0,\quad
e^{\mu|\alpha(2)}h^2_{\mu|\alpha(2)}=0.
$$
It is not difficult to see that from these relations it follows that
$$
h^2_{\mu|\alpha(2)}=0.
$$

Next, let us assume that it is shown that all
$h^k_\mu$, $2\le k\le n_0<N-2$, vanish as a consequence of the equations of
motion. Then the equation of motion
(\ref{9}) at
$n=n_0$ along with the constraint
(\ref{8}) at $n=n_0+2$ give
$$
\varepsilon^{\mu\nu\sigma}e_{\nu|}{}^{\alpha(2)}
h^{n_0+2}_{\sigma|\alpha(n_0+2)}=0,\quad
e^{\mu|}{}_{\alpha(2)}h^{n_0+2}_{\mu|\alpha(n_0+2)}=0,
$$
from where it follows that
$$
h^{n_0+2}_{\mu|\alpha(n_0+2)}=0.
$$
Thus the equation of motion
(\ref{9}) at $n\le N-4$, the constraints (\ref{7}) and
(\ref{8}) along with the equations (\ref{10}) give:
$$
h^n_{\mu|\alpha(n)}=0, \quad n=0,\ldots,N-2.
$$
Finally, the equations of motion
(\ref{9}) at $n=N-2$ and $n=N$ give:
\begin{equation}\label{13}
\varepsilon^{\mu\nu\sigma}e_{\nu|}{}^{\alpha(2)}
h^N_{\sigma|\alpha(N)}=0,
\end{equation}
\begin{equation}\label{14}
\varepsilon^{\mu\nu\sigma}\partial_\nu h^N_{\sigma|\alpha(N)}+
\xi m\varepsilon^{\mu\nu\sigma}e_{\nu|\alpha}{}^\beta
h^N_{\sigma|\alpha(N-1)\beta}=0.
\end{equation}
The condition (\ref{13}) at $N\ge3$ is equivalent to the condition
(\ref{4}). As a consequence of this condition the equations
(\ref{14}) can be re-written in the following equivalent forms:
\begin{equation}\label{15}
\varepsilon_\mu{}^{\nu\sigma}\partial_\nu h^N_{\sigma|\alpha(N)}-
\xi mh^N_{\mu|\alpha(N)}=0,
\end{equation}
$$
e^{\nu|}{}_\beta{}^\gamma\partial_\nu
h^N_{\mu|\alpha(N-1)\gamma}- \xi mh^N_{\mu|\beta\alpha(N-1)}=0.
$$
{}From these equations it follows:
$$
(\Box+m^2)h^N_{\mu|\alpha(N)}=0,\quad
\Box=\partial^2_0-\partial^2_1-\partial^2_2,
$$
\begin{equation}\label{16}
(Wh^N)_{\mu|\alpha(N)}=m\left(\xi{N+2\over2}\right)h^N_{\mu|\alpha(N)}.
\end{equation}
As a result, the parameter
$m$ coincides with the rest mass while from
 (\ref{16}) it follows that the field
$h^N_\mu$ has a spin $\xi(N+2)/2$.

Thus it is shown that the action
(\ref{6}) with the parameters (\ref{12}) describes a
spin
$\xi(N+2)/2$
irreducible
representation of the Poincare group in 2+1 dimensions.

\section{Fermionic Fields}

In this section we derive an action for fermionic fields of an
arbitrary half-integer spin $s$, $|s|=(N + 2)/2\ge3/2$, $N\ge1$ is odd:
$$
S_{F}=
{i\over2}\sum_{n=1}^N\xi_n\varepsilon^{\mu\nu\sigma}h^n_{\mu|}
{}^{\alpha(n)}
\partial_\nu h^n_{\sigma|\alpha(n)}+{i\varepsilon m\over2}
\sum_{n=1}^Na_n
\varepsilon^{\mu\nu\sigma}h^n_{\mu|}{}^{\alpha(n-1)\beta}
e_{\nu|\beta}{}^\gamma h^n_{\sigma|\alpha(n-1)\gamma}+
$$
\begin{equation}\label{17}
im\sum_{n=3}^Nb_n\varepsilon^{\mu\nu\sigma}h^n_{\mu|}{}^{\alpha(n)}
e_{\nu|\alpha(2)}h^{n-2}_{\sigma|\alpha(n-2)}+im\sum_{n=1}^{N-2}
\lambda^{n+2}_{\alpha(n+2)}e^{\mu|\alpha(2)}h^n_{\mu|}{}^{\alpha(n)},
\end{equation}
where summation is carried out over odd
$n$. Fermionic fields $h^n_\mu$ and
$\lambda^n$ are real as well as the coefficients
$a_n$ and $b_n$,
$\xi_n=\pm1$, $\varepsilon=\pm1$.

The constraints have a form:
\begin{equation}\label{18}
e^{\mu|}{}_{\alpha(2)}h^n_{\mu|\alpha(n)}=0, \quad n=1,\ldots,N-2.
\end{equation}
The equations of motion are analogous to the equations of motion for
the bosonic fields:
$$
\xi_n\varepsilon^{\mu\nu\sigma}\partial_\nu h^n_{\sigma|\alpha(n)}+
\varepsilon ma_n\varepsilon^{\mu\nu\sigma}e_{\nu|\alpha}{}^\beta
h^n_{\sigma|\alpha(n-1)\beta}+(1-\delta _{n,1})mb_n\varepsilon^{\mu\nu\sigma}
e_{\nu|\alpha(2)}h^{n-2}_{\sigma|\alpha(n-2)}-
$$
\begin{equation}\label{19}
(1-\delta_{n,N})mb_{n+2}\varepsilon^{\mu\nu\sigma}e_{\nu|}{}^{\alpha(2)}
h^{n+2}_{\sigma|\alpha (n+2)}-(1-\delta_{n,N })me^{\mu|\alpha(2)}
\lambda^{n+2}_{\alpha(n+2)}= 0.
\end{equation}
Analogously to the bosonic case we contract  the equations
(\ref{19}) with
$\partial_\mu$ and exclude the derivatives of all functions
$h^n_\sigma$ with the aid of
(\ref{19}). In the resulting equations  at  fixed $n$
the coefficients in front of
$h^{n+4}_\sigma$  and $h^{n-4}_\sigma$ vanish identically while
the term with
$h^{n-2}_\sigma$ vanishes as a consequence of the constraints
(\ref{18})\footnote{ In fact, the term with
$h^{n-2}_\sigma$ vanishes without using the constraints because
the coefficient in front of $h^{n-2}_\sigma$ vanishes simultaneously
with the coefficient in front of
$h^{n+2}_\sigma$.}. As a result one finds after some transformations:
\begin{equation}\label{20}
\varepsilon b_3\left(\xi_1a_1-{5\over3}\xi_3a_3\right)e^{\sigma|\alpha(2)}
h^3_{\sigma|\alpha(3)}+\left(2\xi_1a^2_1-{4\over3}\xi_3b^2_3\right)
e^{\sigma|}{}_\alpha{}^\beta
h^1_{\sigma|\beta}-{1\over m}e^{\mu|\alpha(2)}\partial_\mu\lambda^3_{\alpha(3)}=0,
\end{equation}
$$
\varepsilon b_{n+2}\left(\xi_na_n-{n+4\over n+2}\xi_{n+2}a_{n+2}\right)
e^{\sigma|\alpha(2)}h^{n+2}_{\sigma|\alpha(n+2)}+
$$
$$
\left({2\over n}\xi_na^2_n+2\xi_{n-2}b^2_n-
{2n(n+3)\over(n+1)(n+2)}\xi_{n+2}b^2_{n+2}\right)e^{\sigma|}{}_\alpha{}^\beta
h^n_{\sigma|\alpha(n-1)\beta}-
$$
\begin{equation}\label{21}
2\xi_{n-2}b_n\lambda^n_{\alpha(n)}-{1\over m}(1-\delta_{n,N})
e^{\mu|\alpha(2)}\partial_\mu\lambda^{n+2}_{\alpha(n+2)}=0, \quad 3\le n\le N.
\end{equation}
Similarly to the bosonic case we define
$b_{N+2}=0$. Let us require that
 the coefficient in front of
$h^3_\sigma$ in the equation
(\ref{20}) and the coefficients in front of
$h^{n+2}_\sigma$ and $h^n_\sigma$ in the equations
 (\ref{21}) vanish:
$$
\xi_na_n-{n+4\over n+2}\xi_{n+2}a_{n+2}=0, \quad 1\le n\le N,
$$
$$
{1\over n}\xi_na^2_n+\xi_{n-2}b^2_n-{n(n+3)\over(n+1)(n+2)}\xi_{n+2}
b^2_{n+2}=0, \quad 3\le n\le N.
$$
These equations are analogous to those in the bosonic case and have the same
solutions. Let us choose a normalization
$a_N=(-1)^{(N+1)/2}$. In addition in the section 5 it will be shown
that the energy boundness conditions along with the requirement
of positivity of norms of states demand
$\xi_N=
(-1)^{(N+1)/2}$.
As a result we have:
\begin{equation}\label{22}
\xi_n=(-1)^{(n+1)/2},\quad a_n=(-1)^{(n+1)/2}{N+2\over n+2},\quad
b_n={(-1)^{(n+1)/2}\over2}\sqrt{{(N+2)^2-n^2\over n(n+1)}}.
\end{equation}

Similarly to the bosonic case, from the equations
(\ref{21}) it follows that all Lagrangian multipliers
$\lambda^n$ vanish while the equation
(\ref{20}) takes a form:
\begin{equation}\label{23}
e^{\mu|}{}_\alpha{}^\beta h^1_{\mu|\beta}=0.
\end{equation}

Let us now turn to the equations for $h^n_\mu$.

Let $N=1$.

For that case, the Lagrangian multipliers are absent while
$h^1_{\mu|\alpha}$ satisfies the equation
$$
\varepsilon^{\mu\nu\sigma}\partial_\nu h^1_{\sigma|\alpha}+
\varepsilon m\varepsilon^{\mu\nu\sigma}e_{\nu|\alpha}{}^\beta
h^1_{\sigma|\beta}=0.
$$
This equation and its consequence
(\ref{23}) are equivalent to the equations
(\ref{4}), (\ref{5}).

Let $N\ge3$.

{}From the constraint
(\ref{18}) at $n=1$ and the equation (\ref{23}) it follows that
$h^1_{\mu|\alpha}=0$. Then, similarly to the bosonic case, one proves that
$h^n_\mu=0$, $3\le n\le N-2$. The equation of motion
(\ref{19}) at $n=N-2$ gives the condition (\ref{13}) for
$h^N_\mu$ which is  equivalent to
(\ref{4}), while the equation of motion (\ref{19}) at $n=N$ gives
\begin{equation}\label{24}
\varepsilon^{\mu\nu\sigma}\partial_\nu h^N_{\sigma|\alpha(N)}+
\varepsilon m\varepsilon^{\mu\nu\sigma}e_{\nu|\alpha}{}^\beta
h^N_{\sigma|\alpha(N-1)\beta}=0,
\end{equation}
from which it follows that
$h^N_\mu$ satisfies the Klein-Gordon equation with the mass
$m$ and describes spin $\varepsilon(N + 2)/2$.

Let us note that if the factors of
$i$ and $\varepsilon$ are absorbed by a redefinition of the coefficients
$\xi_n$, $a_n$ and $b_n$, then the parameters of bosonic and fermionic
actions can be written in a unique form:
$$
\tilde{\xi}_n=\zeta_n\xi,\quad
\tilde{a}_n=\zeta_n\theta{N+2\over n+2},\quad
\tilde{b}_n={\zeta_n\over2}\sqrt{{(N+2)^2-n^2\over n(n+1)}},
$$
$$
\zeta_n=(-1)^{(n+2)/2}=\exp(i\pi{n+2\over2})\,,
$$
where
$$
\tilde{a}_n = i\epsilon a_n ,\qquad
\tilde{\xi}_n = i\xi_n ,\qquad
\tilde{b}_n =i b_n
$$
for the fermionic case and
$$
\tilde{a}_n =  a_n ,\qquad
\tilde{\xi}_n = \xi_n ,\qquad
\tilde{b}_n = b_n
$$
for the bosonic case.
According to the physical arguments explained in the section 5,
$\theta=1$ in the bosonic case and  $\xi=1$ in the fermionic case.

\section{Hamiltonian Formalism}

In this section we consider the limitations on the parameters of the
model under consideration which follow from the requirements of
the positivity of energy and positive definiteness of norms of states.
As shown in the previous sections, the fields
$h^n_\mu$, $n<N$, all Lagrangian multipliers
$\lambda^n$ and some of the components of the field
$h^N_\mu$ vanish as a consequence of the field equations. This implies
that this theory possesses constraints. Since
it is non-degenerate (no gauge symmetry) the constraints can be
only of second class.
As shown in \cite{GT} insertion of second class constraints
into an original action gives raise to some equivalent action.
Let us use this fact. Also, due to the Lorentz invariance, it is
enough for our purpose to consider the components of
$h^N_\mu$ at vanishing momentum (in the rest frame). The equations of motion
(\ref{14}), (\ref{24}) along with the constraints
(\ref{13}) (or, equivalently, (\ref{15})) give at
$\mu=0$:
\begin{equation}
\label{25}
h^N_{0|\alpha(N)}=0.
\end{equation}
Let us introduce a field
$(1/2)e^{\mu|}_{\alpha_1\alpha_2}
h^N_{\mu|\alpha_3\ldots\alpha_{N+2}}$ $\equiv$
$\tilde{h}^{N+2}_{\alpha(N+2)}$, which is symmetric with respect to
all $N+2$ spinorial indices due to the condition
(\ref{13}) (or, equivalently, (\ref{4})). The equation
(\ref{25}) implies that the field
$\tilde{h}^{N+2}$ is traceless (we use a realization of the matrices
$e_\mu$ with $e_{0|\alpha\beta}=\delta_{\alpha\beta})$:
$$
\tilde{h}^{N+2}_{11\alpha(N)}+\tilde{h}^{N+2}_{22\alpha(N)}=0.
$$
Thus, $\tilde{h}^{N+2}$ has only two independent components:
$$
\tilde{h}^{N+2}_{1\ldots1}\equiv2^{-(N +2)/2}q_1,\quad
\tilde{h}^{N+2}_{21\ldots1} \equiv2^{-(N +2)/2}q_2.
$$
Let us express the actions (\ref{6}) and (\ref{17})  in terms of
the variables $q_1$
and
$q_2$
(using the relation $h^N_{\mu|\alpha(N)}$ $=$ $e_{\mu|}{}^{\alpha(2)}
\tilde{h}^{N+2}_{\alpha(N +2)}$).

Bosonic case.

Using $\xi_N=(-1)^{(N+2)/2}\xi$, $a_N=(-1)^{(N+2)/2}\theta$,
after simple transformations one gets
$$
S_{\hbox{B}}=\xi q_2\dot q_1-{\theta m\over2}(q^2_1+q^2_2).
$$
from this expression it follows that
$\xi$ can have an arbitrary sign while
$\theta$ should be positive.

Fermionic case.

Let us set $\xi_N=(-1)^{(N+1)/2}\xi$, $\xi=\pm1$ and $a_N=(-1)^{(N+1)/2}$:
$$
S_{\hbox{F}}={i\xi\over2}(q_1\dot q_1+q_2\dot q_2)-i\varepsilon mq_1q_2.
$$
It is convenient to pass to a new complex variable $\eta$:
$$
\eta={1\over\sqrt2}(q_1+i\xi\varepsilon q_2),\quad q_1={1\over\sqrt2}
(\eta+\eta^\dagger),\quad q_2={\xi\varepsilon\over i\sqrt2}(\eta-\eta^\dagger).
$$
In these terms the action acquires a form:
$$
S_{\hbox{F}}=\xi(i\eta^\dagger\dot\eta-m\eta^\dagger\eta).
$$
from this expression it follows  directly that
$\varepsilon$ can have an arbitrary sign while for
$\xi$ only a value $\xi=1$ is allowed.

\section{Squaring}

In the previous sections we have constructed the actions for the
irreducible representations of the (2+1)--Poincare group, which
can be written in the form
$$
S_{B}={\xi\over2}\varphi D_N\varphi+{1\over2}\varphi
M_N\varphi,
$$
in the bosonic case and
$$
S_{F}={i\over2}\psi D_N\psi+{i\varepsilon\over2}\psi
M_N\psi,
$$
in the fermionic case, where $\varphi$ ($\psi$) denotes
a set of all fields
$h^n_\mu$, $\lambda^n$ in the bosonic (fermionic) case,
$D_N$ is a homogeneous symmetric (antisymmetric)
first order differential operator,
$M_N$ is symmetric (antisymmetric) matrix which does not contain
derivatives. The matrix  $M_N$ is invertible both is the bosonic
and in the fermionic cases (while the operator
$D_N$ is not). It is natural to expect that,
at least in the bosonic case, the operators
$D_N\pm M_N$ are square roots of some second order
differential operator which describes a field with two values
of spin. As mentioned in the
Introduction, such operators arise naturally by a reduction
of the four-dimensional gauge actions.

Here we formulate a bosonic action which is of the second
order in derivatives and describes spin $\pm s$ fields,
such that the second order operator is obtained by squaring of the
first order operator. Let us start with the action
\begin{equation}\label{26}
S={1\over2}\varphi(-D_N+M_N)R(D_N+M_N)\varphi
=-{1\over2}\varphi D_NR D_N\varphi+
{1\over2}\varphi M_NR M_N\varphi,
\end{equation}
where $R$ is some non-degenerate symmetric matrix. The equations of
motion have a form
$$
(-D_N+M_N)R(D_N+M_N)\varphi=0,
$$
provided that the following condition is true:
\begin{equation}\label{27}
M_NR D_N-D_NR M_N=0.
\end{equation}
It is easy to see that for this case the action
(\ref{26}) describes the fields of two spins
$\pm(N+2)/2$. The condition (\ref{27}) can be satisfied by
choosing a matrix $R$ to be of the form
$$
R=M^{-1}_N.
$$
Then the action (\ref{26}) takes a form:
\begin{equation}\label{28}
S=-{1\over2}\varphi D_NM^{-1}_ND_N\varphi+{1\over2}\varphi
M_N\varphi,
\end{equation}
and one can say that the first order operators in the action
(\ref{6}) (and (\ref{17})) are obtained by ``extracting a square
root'' of the second order operator in the action (\ref{28}).

An equivalent action can be obtained by introducing the auxiliary
fields $\omega$ of the same type as $\varphi$:
$$
S^\prime={1\over2}\varphi M_N\varphi+{1\over2}\omega M_N\omega+
\omega D_N\varphi.
$$
Performing in this action a field redefinition:
$$
\varphi={1\over\sqrt2}(\varphi_1+\varphi_2),\quad
\omega={1\over\sqrt2}(\varphi_1-\varphi_2),
$$
one obtains
$$
S^\prime={1\over2}\varphi_1(D_N+M_N)\varphi_1+
{1\over2}\varphi_2(-D_N+M_N)\varphi_2.
$$
Thus the field $\varphi_1$ describes spin $(N+2)/2$, while the field
$\varphi_2$ describes spin $-(N+2)/2$.

Let us note that, in the fermionic case, the action
$$
S=-{\xi\over2}\psi D_NM^{-1}_ND_N\psi+{\xi\over2}\psi
M_N\psi,\quad\xi=\pm1,
$$
which is a counterpart of the bosonic action
(\ref{28}), also describes two spins
$\pm(N+2)/2$. However, for any
$\xi$ it is not acceptable by physical arguments. The simplest
way to see this is to pass to an equivalent action
$$
S^\prime={\xi\over2}\psi M_N\psi+{\xi\over2}\eta M_N\eta+\eta D_N\psi=
{\xi\over2}\psi_1(D_N+M_N)\psi_1+{\xi\over2}\psi_2(-D_N+M_N)\psi_2.
$$
$$
\psi={1\over\sqrt2}(\psi_1+\psi_2),\quad
\eta={1\over\sqrt2}(\psi_1-\psi_2).
$$
It is clear that for any value of
$\xi$ the kinetic term for one of the fermions has a wrong sign.

Let us mention that such a simple decomposition of the action for
two polarizations into a sum of actions for independent components
turns out to be possible due to introducing an adequate set of
auxiliary fields found in this paper. Note that the fields
$\omega$ (more precisely their part associated with the 1-forms
$h^n_{\mu|\alpha(n)}$) can be interpreted as counterparts of the
Lorentz connection in the triadic formulation of gravity.

\section{Concluding Remarks}

An additional reason why the dynamical systems described
in the paper might be of interest is due to their close relationship
with the self-duality equations in 2+2 dimensions
\begin{equation}
\label{sam}
\varepsilon_{\rho\mu}{}^{\nu\sigma}
\partial_\nu h^N_{\sigma|\alpha(N)}=
\partial_\rho h^N_{\mu|\alpha(N)}-
\partial_\mu h^N_{\rho|\alpha(N)}\,,
\end{equation}
which reduce to
(\ref{5}) by virtue of the ``compactifying'' substitution
\begin{equation}
h^N_{3|\alpha(N)}=0\,,\qquad
\partial_3 h^N_{\mu|\alpha(N)}= \pm m
h^N_{\mu|\alpha(N)}\,,
\end{equation}
under condition that 3--components of vectors correspond to
a second time-like direction. Similarly, the Euclidian version of
(\ref{5}) can be obtained from the Euclidian version of
(\ref{sam}).

Let us note that the set of fields used in the present paper
$h^n_{\mu|\alpha(n)}$ ($0\leq n\leq N$) corresponds to the decomposition
of the fields
$h^N_{\mu|\alpha({N/2})\dot{\alpha}({N/2})}$
($N$ even) and
$h^N_{\mu|\alpha({(N+1)/2})\dot{\alpha}({(N-1)/2})}$
($N$- odd), which were used in
\cite{v1} for the description of four-dimensional massless
fields of spin
$s=N/2 +1$, into irreducible representations of the
diagonal (2+1)--Lorentz group acting both on
 dotted and undotted spinor indices.
Also it is worth mentioning that the Lagrange multipliers
$\lambda^n_{\mu|\alpha(n)}$ have a similar structure in
spinor indices ($N\geq n >0$). This suggests an idea that
one can try to identify them with a component of the 3+1-
dimensional gauge field along  the ``extra'' time within
the dimensional reduction of some hypothetical self-dual theory
in 2+2 - dimensional space-time.

Self-dual equations of motion of
$D=4$ massless fields of an arbitrary spin can be formulated
not only on the linearized level but on the non-linear level
too \cite{v2}. Recently, supersymmetric
self--dual equations for the fields
of arbitrary spin have been studied in  \cite{cho}.
In the context of results obtained in this paper it would be
interesting to come back to the question of a possibility
of formulation of a theory of self-dual fields of an arbitrary spin
at the action level.

Also it would be interesting to find a topologically massive
version of the 2+1 - dimensional theory of massive fields of an
arbitrary spin thus  generalizing the previously constructed theories
for $s=1$ \cite{TOP1,DJT}, $s=3/2$ \cite{DK}, $s=2$ \cite{DJT} and $s=5/2$
\cite{AD}.

\vskip0.3cm
\noindent
The authors are grateful to S.F.~Prokushkin for the collaboration at
the early stage of the work related to the analysis of spin 3.
The research described in this report
 was supported in part by the
Russian Foundation for Basic Research, Grant
 {\sl N} 96-02-17314 and
by the European Community
Commission under the contract INTAS,
Grant {\sl N} 94-2317-{\it ext} and Grant of the Dutch NWO Organization.

\end{document}